\documentclass[journal]{IEEEtran}
\usepackage{amsfonts}
\usepackage{amsmath}
\usepackage{amssymb}
\usepackage{color}
\usepackage{epsfig}
\usepackage{graphics}
\usepackage{graphicx}

\epsfclipon
\newtheorem{theorem}{Theorem}[section]

\newtheorem{corollary}[theorem]{Corollary}

\newtheorem{example}{Example}[section]

\newtheorem{lemma}[theorem]{Lemma}

\newtheorem{problem}{Problem}
\newtheorem{proposition}{Proposition}
\newtheorem{remark}{Remark}[section]

\begin{document}

\title{Moment-Based Spectral Analysis of Large-Scale Networks Using Local
Structural Information}
\author{Victor M. Preciado,~\IEEEmembership{Member,~IEEE,} and Ali
Jadbabaie,~\IEEEmembership{Senior~Member,~IEEE} \thanks{%
Manuscript Received ---------.} \thanks{%
The authors are with the Department of Electrical and Systems Engineering at
the University of Pennsylvania, Philadelphia, PA 19104 USA. (e-mail:
preciado@seas.upenn.edu; jadbabai@seas.upenn.edu).} \thanks{%
This work was supported by ONR MURI \textquotedblleft Next Generation
Network Science\textquotedblright\ and AFOSR \textquotedblleft Topological
And Geometric Tools For Analysis Of Complex Networks\textquotedblright.}}
\maketitle

\begin{abstract}
The eigenvalues of matrices representing the structure of large-scale
complex networks present a wide range of applications, from the analysis of
dynamical processes taking place in the network to spectral techniques
aiming to rank the importance of nodes in the network. A common approach to
study the relationship between the structure of a network and its
eigenvalues is to use synthetic random networks in which structural
properties of interest, such as degree distributions, are prescribed.
Although very common, synthetic models present two major flaws: (\emph{i})
These models are only suitable to study a very limited range of structural
properties, and (\emph{ii}) they implicitly induce structural properties
that are not directly controlled and can deceivingly influence the network
eigenvalue spectrum. In this paper, we propose an alternative approach to
overcome these limitations. Our approach is not based on synthetic models,
instead, we use algebraic graph theory and convex optimization to study how
structural properties influence the spectrum of eigenvalues of the network.
Using our approach, we can compute with low computational overhead
global spectral properties of a network from its local structural
properties. We illustrate our approach by studying how structural
properties of online social networks influence their eigenvalue spectra.
\end{abstract}

\section{Introduction}

During the last decade, the complex structure of many large-scale networked
systems has attracted the attention of the scientific community \cite{S01}.
The availability of massive databases describing these networks allows
researchers to explore their structural properties with great detail.
Statistical analysis of empirical data has unveiled the existence of
multiple common patterns in a large variety of network properties, such as
power-law degree distributions \cite{BA99}, or the small-world phenomenon 
\cite{WS98}. Aiming to replicate these structural patterns, a variety of
synthetic network models has been proposed in the literature, such as the
classical Erd\"{o}s-R\'{e}nyi random graph (and its generalizations) \cite%
{NSW01,New09}, the preferential attachment model proposed by Barab%
\'{a}si and Albert \cite{BA99}, or the small-world network proposed by Watts
and Strogatz \cite{WS98}.


Synthetic network models have been widely used to analyze the performance of
dynamical processes on a network. In this direction, a fundamental question
is to understand the impact of a particular structural property in the
performance of the network \cite{BLMCH06}. The most common approach to
address this question is to use synthetic network models in which one can
prescribe the structural property under study. The impact of structural
features, such as degree distributions \cite{NSW01}, or clustering \cite{New09}, has been widely studied in the literature
following this methodology. Although very common in the literature, this
approach presents two major flaws:

\begin{enumerate}
\item Synthetic network models are only suitable to study a very limited
range of structural properties. For example, synthetic random networks
presenting structural properties beyond simple degree distributions become
intractable from a spectral point of view.

\item Synthetic network models implicitly induce many structural properties
that are not directly controlled and can be relevant to the network
dynamical performance. Therefore, it is difficult to isolate
the role of a particular structural property using synthetic network models.
\end{enumerate}

Since a network's eigenvalues influence the dynamical behavior of dynamical
processes that can take place in the network \cite{WCWF03}-\cite{Pre08}, it
is of interest to study the relationship between the structural properties
of the network, such as the distribution of degrees, triangles and other
substructures, and its eigenvalue spectrum. In this paper, we propose a
novel framework, based on spectral and algebraic graph theory and convex
optimization, to compute with low computational overhead global spectral
properties of a network from its local structural properties. In
particular, we derive optimal bounds and estimators of spectral properties
of interest from structural information. Our results are useful to unveil
the set of structural properties that have the highest impact in the
eigenvalue spectrum of a network. In particular, in the case of online
social networks, we find that the correlation between the distribution of
degrees and triangles in the network plays a key role in the spectral radius.

The rest of this paper is organized as follows. In the next section, we
review graph-theoretical terminology needed in our derivations. We also
review existing bounds and estimators of spectral properties of a network in
terms of structural properties. In Section~\ref{Moments from Metrics}, we
use algebraic graph theory to derive closed-form expressions for the
so-called spectral moments of a network. In Section \ref{Optimal Bounds}, we
use convex optimization to derive optimal bounds on spectral properties of
interest from these moments. We numerically verify the performance of our
bounds using real network data in Section \ref{Simulations}, where we also
use our results to unveil the set of structural properties with the highest
influence on the spectral radius of social networks.

\section{\label{Notation}Notation \& Preliminaries}

Let $\mathcal{G}=\left( \mathcal{V},\mathcal{E}\right) $ denote an
undirected graph with $n$ nodes, $e$ edges, and no self-loops\footnote{%
An undirected graph with no self-loops is also called a \emph{simple} graph.}%
. We denote by $\mathcal{V}\left( \mathcal{G}\right) =\left\{ v_{1},\dots
,v_{n}\right\} $ the set of nodes and by $\mathcal{E}\left( \mathcal{G}%
\right) \subseteq\mathcal{V}\left( \mathcal{G}\right) \times\mathcal{V}%
\left( \mathcal{G}\right) $ the set of undirected edges of $\mathcal{G}$. If 
$\left\{ v_{i},v_{j}\right\} \in\mathcal{E}\left( \mathcal{G}\right) $ we
call nodes $v_{i}$ and $v_{j}$ \emph{adjacent} (or neighbors), which we
denote by $v_{i}\sim v_{j}$. We define a \emph{walk} of length $k$ from $%
v_{0}$ to $v_{k}$ to be an ordered sequence of nodes $\left(
v_{0},v_{1},...,v_{k}\right) $ such that $v_{i}\sim v_{i+1}$ for $%
i=0,1,...,k-1$. If $v_{0}=v_{k}$, then the walk is closed. A closed walk
with no repeated nodes (with the exception of the first and last nodes) is
called a \emph{cycle}. For example, \emph{triangles}, \emph{quadrangles} and 
\emph{pentagons} are cycles of length three, four, and five, respectively.

Graphs can be algebraically represented via matrices. The adjacency matrix
of an undirected graph $\mathcal{G}$, denoted by $A_{\mathcal{G}}=[a_{ij}]$,
is an $n\times n$ symmetric matrix defined entry-wise as $a_{ij}=1$ if nodes 
$v_{i}$ and $v_{j}$ are adjacent, and $a_{ij}=0$ otherwise\footnote{%
For simple graphs, $a_{ii}=0$ for all $i$.}. The eigenvalues of $A_{\mathcal{%
G}}$, denoted by $\lambda _{1}\geq \lambda _{2}\geq \ldots \geq \lambda _{n}$%
, play a key role in our paper. The spectral radius of $A_{\mathcal{G}}$,
denoted by $\rho \left( A_{\mathcal{G}}\right) $, is the maximum among the
magnitudes of its eigenvalues. Since $A_{\mathcal{G}}$ is a symmetric matrix
with nonnegative entries, all its eigenvalues are real and the spectral
radius is equal to the largest eigenvalue, $\lambda _{1}$. We define the $k$%
-th spectral moment of the adjacency matrix $A_{\mathcal{G}}$ as%
\begin{equation}
m_{k}\left( A_{\mathcal{G}}\right) =\frac{1}{n}\sum_{i=1}^{n}\lambda
_{i}^{k}.  \label{Spectral Moment}
\end{equation}%
As we shall show in Section \ref{Moments from Metrics}, there is a direct
connection between the spectral moments and the presence of certain
substructures in the graph, such as cycles of length $k$.

We define the set of neighbors of $v$ as $\mathcal{N}_{v}=\{w\in \mathcal{V}%
\left( \mathcal{G}\right) :\left\{ v,w\right\} \in \mathcal{E}\left( 
\mathcal{G}\right) \}$. The number of neighbors of $v$ is called the \emph{%
degree} of node $v$, denoted by $d_{v}$. We can define several local
neighborhoods around a node $v$ based on the concept of distance. Let $%
d\left( v,w\right) $ denote the \emph{distance} between two nodes $v$ and $w$
(i.e., the minimum length of a walk from $v$ to $w$). We say that $v$ and $w$
are $k$-hop neighbors if $d\left( v,w\right) =k,$ and define the $k$-th
order neighborhood of $v$ as $\mathcal{N}_{v}^{\left( k\right) }=\left\{ w\in%
\mathcal{V}\left( \mathcal{G}\right) :d\left( v,w\right) \leq k\right\} $.
The set of nodes in $\mathcal{N}_{v}^{\left( k\right) }$ induces a subgraph $%
\mathcal{G}_{v}^{\left( k\right) }\subseteq\mathcal{G}$, with node-set $%
\mathcal{N}_{v}^{\left( k\right) }$ and edge-set $\mathcal{E}_{v}^{\left(
k\right) } \subseteq \mathcal{E}\left( \mathcal{G}\right)$ defined as the
subset of edges connecting nodes in $\mathcal{N}_{v}^{\left( k\right) }$.

\subsection{Estimators of the Spectral Radius}

Random network models are currently the primary tool to study the
relationship between the structure and dynamics of complex networks \cite%
{BLMCH06}. Although many random networks have been proposed to analyze
structural properties such as the degree distribution \cite{NSW01}, or clustering \cite{New09}, only random networks including a very limited amount of structural
information are currently amenable to spectral analysis.

In the original Erd\"{o}s-R\'{e}nyi random graph with $n$ nodes, denoted by $%
G\left( n,p\right)$, each edge is independently chosen with a fixed
probability $p$, \cite{ER61}. In this model, all the nodes present the same
expected degree, $\mathbb{E}[d_{i}]=np$, and the largest eigenvalue of its
adjacency matrix is almost surely $\left[ 1+o\left( 1\right) \right] np$
(assuming that $np=\Omega \left( \log n\right) $). Although very interesting
from a theoretical point of view, the original random graph presents very
limited modeling capabilities, since the degree distributions of real-world
networks are almost never uniform.

In order to increase the modeling abilities of random graphs, several models
have been proposed in the literature. For example, given a sequence $\mathbf{%
w}=\left( w_{1},...,w_{n}\right) $, Chung and Lu proposed in \cite{CL03} a
random graph $G\left( \mathbf{w}\right) $ with an expected sequence of
degrees equal to $\mathbf{w}$. In this random graph, edges are independently
assigned to each pair of vertices $\left( i,j\right) $ with probability $%
\left. w_{i}w_{j}\right/ \sum_{k=1}^{n}w_{k}$. Chung et al.
proved in \cite{CLV03} that if $\left. \sum_{i=1}^{n}w_{i}^{2}\right/
\sum_{j=1}^{n}w_{j}>\sqrt{\max \left\{ w_{i}\right\} }\log n$, then the
largest eigenvalue $\lambda _{1}\left( G\left( \mathbf{w}\right) \right) $
converges almost surely%
\begin{equation}
\lambda _{1}\left( G\left( \mathbf{w}\right) \right) \overset{a.s.}{%
\rightarrow }\left[ 1+o\left( 1\right) \right] \frac{\sum_{i=1}^{n}w_{i}^{2}%
}{\sum_{j=1}^{n}w_{j}},  \label{CL Estimator}
\end{equation}%
for large $n$. Despite its theoretical interest, random graphs with a given
degree distribution are by far not enough to faithfully model the structure
of real complex networks.





Although random graph models with more elaborated structural properties,
such as clustering or hierarchy, can be found in the literature, these
models are usually extremely challenging (if not impossible) to analyze from
a spectral point of view. The source of this intractability is the presence
of strong correlations among the entries of the (random) adjacency matrices
associated with these models. In Section \ref{Moments from Metrics} and \ref%
{Optimal Bounds}, we introduce an alternative method to analyze the effect
of elaborated structural properties, such as clustering and correlations, on
the eigenvalues of a network without the use of intractable random graphs
models.


\subsection{\label{Viral Bounds}Bounds on the Spectral Radius}

In this subsection, we review some existing bounds relating structural
features of a network, such as the degree distribution, with the spectral
radius of the network. We can find in the literature several bounds on the
spectral radius that are not based on random models. For example, for a
graph $\mathcal{G}$ with $n$ nodes and $e$ edges, we have the following
upper bounds for the spectral radius \cite{ZV08}:%
\begin{align*}
u_{1}& =\sqrt{2e-\left( n-1\right) d_{\min }+\left( d_{\min }-1\right)
d_{\max }}, \\
u_{2}& =\max \left\{ \sqrt{d_{i}m_{j}}\text{, }\left( i,j\right) \in
E\right\},
\end{align*}%
where $d_{\min }$ and $d_{\max }$ are the minimum and maximum degrees of $%
\mathcal{G}$, and $m_{i}=\frac{1}{d_{i}}\sum_{j\in \mathcal{N}_{i}}d_{j}$.
Notice that none of the above bounds take into account the presence of
triangles, or other cycles, in the graph. Since many real-world networks
present a high density of cycles (i.e., social graphs), these bounds perform
poorly in many real applications. In the following sections, we propose a
methodology to derive bounds on spectral properties of relevance in terms of
a wide variety of structural features, including the distribution of cycles
in the network.

\section{\label{Moments from Metrics}Moment-Based Analysis of the Adjacency
Spectrum}

Algebraic graph theory provides us with tools to relate the eigenvalues of a
network with its structural properties. Particularly useful is the following
result relating the $k$-th spectral moment of $A_{\mathcal{G}}$ with the
number of closed walks of length $k$ in $\mathcal{G}$ \cite{Big93}:

\medskip

\begin{lemma}
\label{Moments from Walks}Let $\mathcal{G}$ be a simple graph. The $k$-th
spectral moment of the adjacency matrix of $\mathcal{G}$ can be written as%
\begin{equation}
m_{k}(A_{\mathcal{G}})=\frac{1}{n}\sum_{i=1}^{n}\lambda_{i}^{k}=\frac{1}{n}%
\left\vert \Psi_{\mathcal{G}}^{\left( k\right) }\right\vert ,
\label{Moments as Walks in Graph}
\end{equation}
where $\Psi_{\mathcal{G}}^{\left( k\right) }$ is the set of all closed walks
of length $k$ in $\mathcal{G}$. \footnote{%
We denote by $\left\vert Z\right\vert $ the cardinality of a set $Z$.}
\end{lemma}

\medskip

In the following subsections, we build on Lemma \ref{Moments from Walks} to
compute the spectral moments of a network in terms of relevant structural
features.

\subsection{Low-Order Spectral Moments}

From (\ref{Moments as Walks in Graph}), we can easily compute the first
three moments of $A_{\mathcal{G}}$ in terms of the distribution of degrees
and triangles as follows \cite{Big93}:

\medskip

\begin{corollary}
\label{One corollary}Let $\mathcal{G}$ be a simple graph with adjacency
matrix $A_{\mathcal{G}}$. Denote by $d_{i}$ and $t_{i}$ the number of edges
and triangles touching node $i\in\mathcal{V}\left( \mathcal{G}\right) $,
respectively. Then,%
\begin{align}
m_{1}(A_{\mathcal{G}}) & =0,  \label{algebraic graph} \\
m_{2}(A_{\mathcal{G}}) & =\frac{1}{n}\sum_{i\in\mathcal{V}\left( \mathcal{G}%
\right) }d_{i},  \notag \\
m_{3}(A_{\mathcal{G}}) & =\frac{1}{n}\sum_{i\in\mathcal{V}\left( \mathcal{G}%
\right) }2t_{i}.  \notag
\end{align}
\end{corollary}

\medskip

\begin{proof}
Since there are no self-loops in a simple graph, we have that $m_{1}(A_{%
\mathcal{G}})=0$. In order to compute $m_{2}(A_{\mathcal{G}})$, we need to
count the number of closed walks of length two starting at a node $i$. The
number of walks of this type is equal to $d_{i}$. Summing over all possible
starting nodes we obtain $\sum_{i\in \mathcal{V}\left( \mathcal{G}\right)
}d_{i}$. The third moment is proportional to the number of closed walks of
length 3. Starting at node $i$, there are $2t_{i}$ walks of this type, where
the coefficient 2 accounts for the two possible directions one can walk each
triangle. Summing over all possible starting points, we obtain $\sum_{i\in 
\mathcal{V}\left( \mathcal{G}\right) }2t_{i}$.
\end{proof}

\medskip These moments can also be expressed in terms of the total number of
edges and triangles in $\mathcal{G}$, which we denote by $e$ and $\Delta $,
respectively. Since $e=\frac{1}{2}\sum_{i}d_{i}$ and $\Delta =\frac{1}{3}%
\sum_{i}t_{i}$ \cite{Big93}, we have that: 
\begin{align}
m_{1}(A_{\mathcal{G}})& =0,  \label{Moments as Averages} \\
m_{2}(A_{\mathcal{G}})& =2e/n,  \notag \\
m_{3}(A_{\mathcal{G}})& =6\,\Delta /n.  \notag
\end{align}%
where the coefficients 2 (resp. 6) in the above expressions corresponds to
the number of closed walks of length 2 (resp. 3) enabled by the presence of
an edge (resp. triangle). The computation of higher-order moments requires a
more elaborated combinatorial analysis. We include details for the fourth
and fifth spectral moments in the following subsections.

\subsection{Fourth-Order Spectral Moments}

\begin{figure*}[t]
\centering
\includegraphics[width=0.85\textwidth]{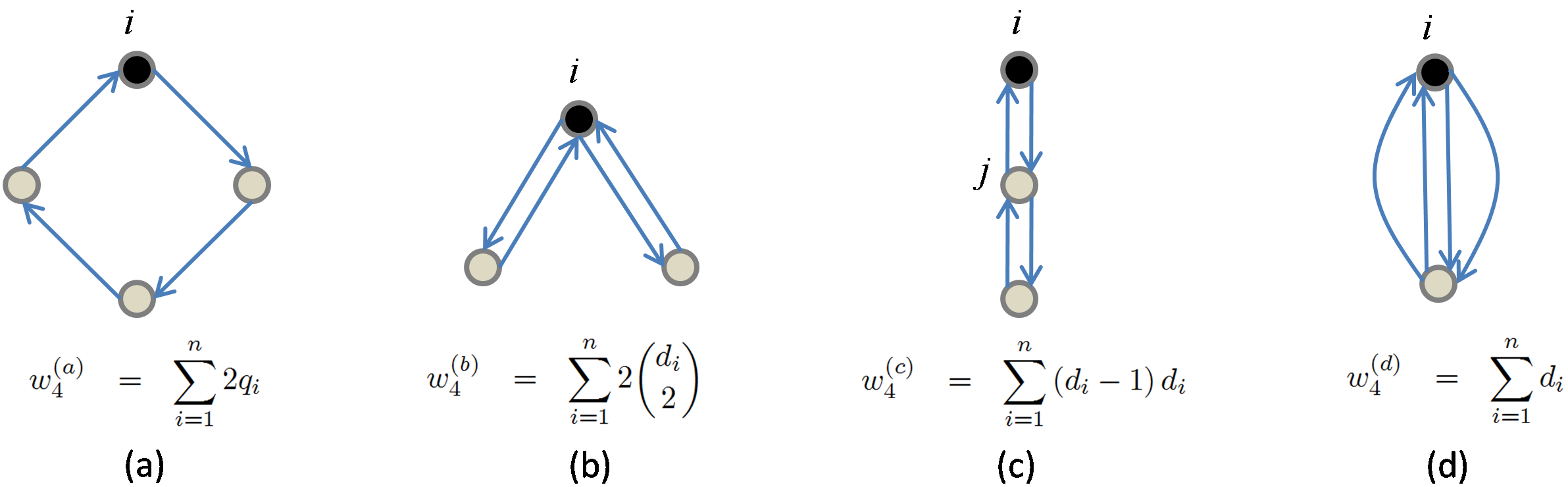}
\caption{Enumeration of the possible types of closed walks of length $4$ in
a graph with no self-loops. The classification is based on the structure of
the subgraph underlying each closed walk. For each walk type, we also
include an expression that corresponds to the number of closed walks of that
particular type in terms of network structural properties.}
\label{fig_4}
\end{figure*}

A combinatorial analysis of (\ref{Moments as Walks in Graph}) for $k=4$
gives us the following result:

\medskip

\begin{lemma}
\label{Lemma 4th metrics}Let $\mathcal{G}$ be a simple graph with adjacency
matrix $A_{\mathcal{G}}$. Denote by $q_{i}$ and $d_{i}$ the number of
quadrangles and edges touching node $i\in\mathcal{V}\left( \mathcal{G}%
\right) $, respectively. Then,%
\begin{equation}
m_{4}\left( A_{\mathcal{G}}\right) =\frac{1}{n} \sum_{i\in \mathcal{V}\left( 
\mathcal{G}\right) }2q_{i}+4\binom{d_{i}}{2}+d_{i} .
\label{Fourth Moment Metrics}
\end{equation}
\end{lemma}

\medskip

\begin{proof}
We compute the fourth moment from (\ref{Moments as Walks in Graph}) by
counting the number of closed walks of length 4 in $\mathcal{G}$. In Fig. 1,
we enumerate all the possible types of closed walks of length $4$. We can
count the number of closed walks of each particular type in terms of network
structural features as follows:

\begin{description}
\item[(a)] The number of closed walks of type (a) is equal to twice the
number of quadrangles, where the coefficient $2$ in $w_{4}^{\left( a\right)
} $ accounts for the two possible directions (clockwise and
counterclockwise) one can walk each quadrangle.

\item[(b)] The number of walks of type (b) starting at node $i$ is equal to $%
2\binom{d_{i}}{2}$. The expression for $w_{4}^{\left( b\right) }$ comes from
summing over all possible starting points, $i=1,...,n.$

\item[(c)] The number of closed walks of this type can also be written in
terms of the degrees as:%
\begin{equation*}
w_{4}^{\left( c\right) }=\sum_{i=1}^{n}\sum_{j=1}^{n}a_{ij}\left(
d_{j}-1\right) =\sum_{j=1}^{n}\left( d_{j}-1\right) d_{j}.
\end{equation*}

\item[(d)] The number of closed walks of this type starting at node $i$ is
equal to $d_{i}$, thus, $w_{4}^{\left( d\right) }=\sum_{i=1}^{n}d_{i}$.
\end{description}

Hence, we obtain (\ref{Fourth Moment Metrics}) by summing up all the above
contributions, $w_{4}^{\left( a\right) }+w_{4}^{\left( b\right)
}+w_{4}^{\left( c\right) }+w_{4}^{\left( d\right) }$, and simple algebraic
manipulations).
\end{proof}

\medskip Lemma \ref{Lemma 4th metrics} provides an expression to compute the
fourth spectral moment in terms of structural features, namely, the
distribution of degrees and quadrangles. We illustrate Lemma \ref{Lemma 4th
metrics} in the following example. \medskip

\begin{example}
\label{Ring graph}Consider the $n$-ring graph, $R_{n}$ (without self-loops).
The eigenvalues of the adjacency matrix of the ring graph are $\lambda
_{i}\left( A_{R_{n}}\right) =2\cos i\frac{2\pi }{n}$, for $i=0,1,...,n-1$.
Hence, the fourth moment is equal to $m_{4}(A_{R_{n}})=\frac{1}{n}%
\sum_{i=0}^{n-2}\left( 2\cos i\frac{2\pi }{n}\right) ^{4}$, which (after
some computations) can be found to be equal to $6$ for $n\not\in \left\{
2,4\right\} $. We can reach this same result by directly applying (\ref%
{Fourth Moment Metrics}), without performing an eigenvalue decomposition, as
follows. In the ring graph, we have that $d_{i}=2$ and $q_{i}=0$, for $%
n\not\in \left\{ 2,4\right\} $. Hence, from (\ref{Fourth Moment Metrics}),
we directly obtain $m_{4}(A_{R_{n}})=6$, for $n\not\in \left\{ 2,4\right\} $.
\end{example}

\medskip

The fourth spectral moment\ can be rewritten in terms of aggregated
quantities, such as the total number of quadrangles and edges, and the
sum-of-squares of the degrees, as follows:

\medskip

\begin{corollary}
\label{Cor Fourth Subgraphs}Let $\mathcal{G}$ be a simple graph. Denote by $%
e $ and $Q$ the total number of edges and quadrangle in $\mathcal{G}$,
respectively, and define $W_{2}=\sum_{i=1}^{n}d_{i}^{2}$. Then,%
\begin{equation}
m_{4}\left( A_{\mathcal{G}}\right) =\frac{1}{n}\left[ 8Q+2W_{2}-2e\right] .
\label{Fourth Moment Subgraphs}
\end{equation}
\end{corollary}

\medskip

\begin{proof}
The proof comes straightforward from (\ref{Fourth Moment Metrics}) by
substituting $\sum_{i=1}^{n}q_{i}=4Q$ and $\sum_{i=1}^{n}d_{i}=2e.$
\end{proof}

\medskip

Hence, we do not need to have access to the detailed distribution of
quadrangles and degrees in $\mathcal{G}$ to compute the fourth moment, we
only need to know the aggregated quantities $Q$, $W_{2}$, and $e$.

\subsection{Fifth-Order Moment}

\begin{figure*}[t]
\centering
\includegraphics[width=1.0\textwidth]{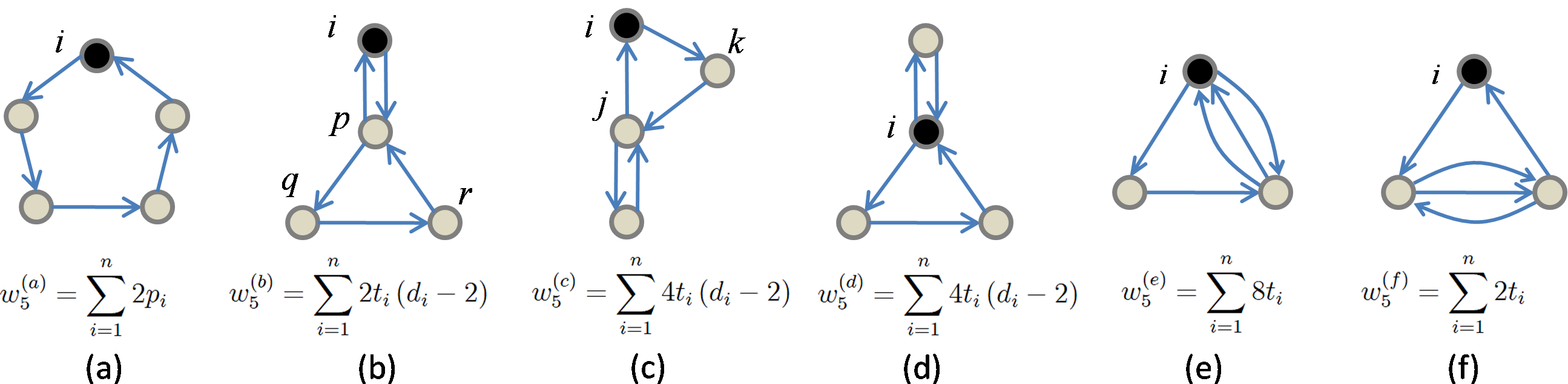}
\caption{Possible types of closed walks of length $5$ in a simple graph. The
classification is based on the structure of the subgraph underlying the
closed walks. For each walk type, we also include an expression that
corresponds to the number of closed walks of that particular type in terms
of network structural features.}
\label{fig_5}
\end{figure*}

\begin{lemma}
\label{Lemma 5th metrics}Let $\mathcal{G}$ be a simple graph. Denote by $%
p_{i}$, $t_{i}$, and $d_{i}$ the number of pentagons, triangles, and edges
touching node $i\in\mathcal{V}\left( \mathcal{G}\right) $, respectively.
Then,%
\begin{equation}
m_{5}\left( A_{\mathcal{G}}\right) =\frac{1}{n} \sum_{i\in \mathcal{V}\left( 
\mathcal{G}\right) }2p_{i}+10t_{i}d_{i}-10t_{i} .
\label{Fifth Moment Metrics}
\end{equation}
\end{lemma}

\medspace

\begin{proof}
The proof follows the same structure as that of Lemma \ref{Lemma 4th metrics}%
. A graphical representation of the types of closed walks of length 5 is
provided in Fig. 2. Details regarding the counting of closed walks of each
particular type can be found in the Appendix.
\end{proof}

\medspace

Lemma \ref{Lemma 5th metrics} expresses the fifth spectral moment of $A_{%
\mathcal{G}}$ in terms of network structural features. We can rewrite (\ref%
{Fifth Moment Metrics}) in terms of aggregated quantities as follows:

\medspace

\begin{corollary}
\label{Cor Fifth Moment}Let $\mathcal{G}$ be a simple graph. Denote by $%
\Delta$ and $\Pi$ the total number of triangles and pentagons in $\mathcal{G}
$, respectively. Define the degree-triangle correlation as $\mathcal{C}%
_{dt}=\sum_{i}d_{i}t_{i}$. Then,%
\begin{equation}
m_{5}\left( A_{\mathcal{G}}\right) =\frac{1}{n}\left[ 10\Pi+10\mathcal{C}%
_{dt}-30\Delta\right] .  \label{Fifth Moment Subgraphs}
\end{equation}
\end{corollary}

\medspace

\begin{proof}
The proof comes from (\ref{Fifth Moment Metrics}) taking into account that $%
\sum_{i=1}^{n}p_{i}=5\Pi$ and $\sum_{i=1}^{n}t_{i}=3\Delta.$
\end{proof}

\medspace

Observe how, as we increase the order of the moments, more complicated
structural features appear in the expressions. In particular, the sum and
sum-of-squares of the degrees influence the second and fourth spectral
moments. (Notice that we can expand $\binom{d_{i}}{2}=\frac{1}{2}\left(
d_{i}^{2}-d_{i}\right) $ in (\ref{Fourth Moment Metrics})). The total number
of triangles in the network, $\Delta $, influences the third and fifth
moments in (\ref{Moments as Averages}) and (\ref{Fifth Moment Subgraphs}).
Also, the correlation between degree and triangle distributions, quantified
by $\mathcal{C}_{dt}=\sum_{i}d_{i}t_{i}$, influences the fifth spectral
moment in (\ref{Fifth Moment Subgraphs}). We shall show in Section \ref%
{Simulations} that this structural correlation strongly influences the
spectral radius of online social networks.



The main advantage of our results may not be apparent in networks with
simple, regular structure. For these networks, an explicit eigenvalue
decomposition is usually easy to compute and there may be no need to look
for alternative ways to compute spectral properties. On the other hand, in
the case of large-scale complex networks, the structure of the network can
be very intricate ---in many cases not even known exactly--- and an explicit
eigenvalue decomposition can be very challenging to compute, if not
impossible. It is in these cases when the alternative approach proposed in
this paper is most useful. In the following subsection, we use our
expressions to compute the spectral moments of an online social network from
empirical structural data.

\subsection{Spectral Moments of an Online Social Network\label{Social Graph
Moments}}

The real network under study is a subgraph of Facebook with $2,404$ nodes
and $22,786$ edges obtained from crawling the graph in a breadth-first
search around a particular node (the dataset can be found in \cite{MAT1}).
Although the approach proposed in this section is meant to be used for much
larger networks, we illustrate our results with this medium-size subgraph in
order to compare our analysis with the results obtained from an explicit
eigenvalue decomposition of the complete network topology.

In this example, we first compute the structural metrics involved in the
first five spectral moments, in particular, the degree $d_{i}$, the number
of triangles $t_{i}$, quadrangles $q_{i}$, and pentagons $p_{i}$ touching
each node $i\in \mathcal{V}$. The degree $d_{i}$ and the number of triangles 
$t_{i}$ touching node $i$ can be easily computed by counting the number of
edges attached to node $i$ and the number of edges connecting friends of $i$%
, respectively. In order to count the number of quadrangles $q_{i}$ and
pentagons $p_{i}$ touching each node $i$, we must know the structure of the
network around node $i$ with a radius of 2, i.e., node $i$ needs to know who
her friends' friends are. We denote by $\left\vert N_{i,2}\right\vert $ the
number of nodes in the two-hops neighborhood around node $i$ (excluding node 
$i$). In order to count the number of quadrangles (resp. pentagons) touching
node $i$, we must verify the presence of a cycle for each one of the $\binom{%
\left\vert N_{i,2}\right\vert }{3}$ (resp. $\binom{\left\vert
N_{i,2}\right\vert }{4}$) subsets of three (resp. four) nodes in $N_{i,2}$.

In Fig. 3, we plot the distributions of degrees and triangles, as well as a
scatter plot of $t_{i}$ versus $d_{i}$ (where each point has coordinates $%
\left( d_{i},t_{i}\right) $, in log-log scale, for all $i\in \mathcal{V}%
\left( \mathcal{G}\right) $). We then aggregate, via simple averaging, those
structural metrics that are relevant to compute the spectral moments. In
particular, we obtain the following numerical values for these metrics:%
\begin{equation*}
\begin{array}{ll}
e/n=\sum d_{i}/2n=9.478, & \Delta /n=\sum t_{i}/3n=28.15, \\ 
Q/n=\sum q_{i}/4n=825.3, & \Pi /n=\sum p_{i}/5n=31,794, \\ 
W_{2}/n=\sum d_{i}^{2}/n=1,318, & \mathcal{C}_{dt}/n=\sum d_{i}t_{i}/n=8,520.%
\end{array}%
\end{equation*}%
Hence, using these values in expressions (\ref{algebraic graph}), (\ref%
{Fourth Moment Subgraphs}), and (\ref{Fifth Moment Subgraphs}), we obtain
the following spectral moments: $m_{1}\left( A_{\mathcal{G}}\right) =0,$ $%
m_{2}\left( A_{\mathcal{G}}\right) =18.95,$ $m_{3}\left( A_{\mathcal{G}%
}\right) =168.9,$ $m_{4}\left( A_{\mathcal{G}}\right) =9,230,$ and $%
m_{5}\left( A_{\mathcal{G}}\right) =402,310.$


In this section, we have derived expressions to compute the first five
spectral moment of $A_{\mathcal{G}}$ from network structural properties. In
the next section, we use semidefinite programming to extract bounds on
spectral properties of interest from a sequence of spectral moments.

\medspace
\begin{figure*}[t]
\centering
\includegraphics[width=1.0\textwidth]{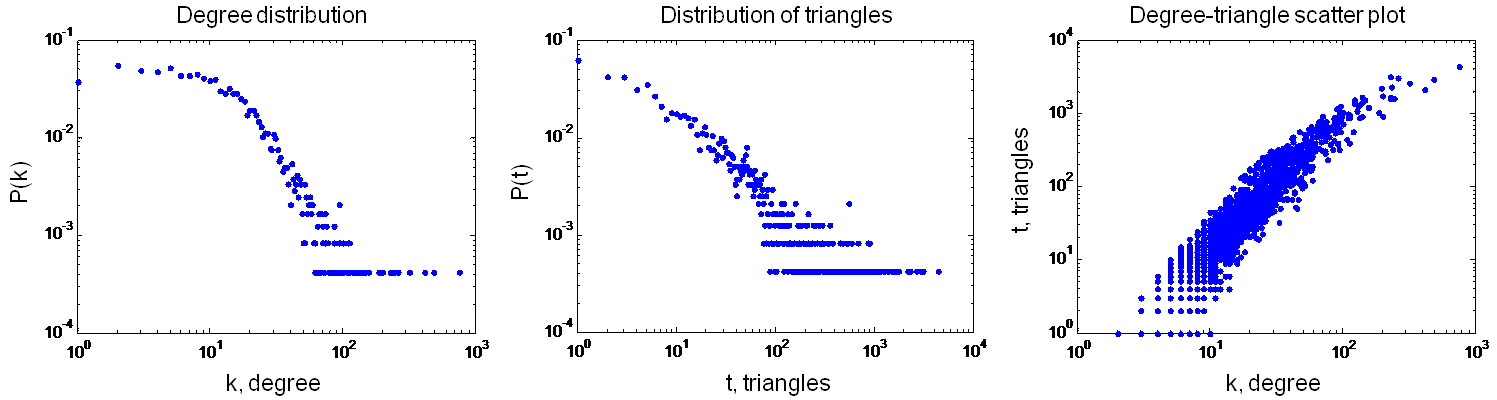}
\caption{In the left and center figures, we plot the distributions of
degrees and triangles of the social network under study (in log-log scale).
In the right figure, we include a scatter plot where each point has
coordinates $(d_{i},t_{i})$, in log-log scale, for all the nodes in the
social graph.}
\label{fig_7}
\end{figure*}

\section{\label{Optimal Bounds}Optimal Spectral Bounds from Spectral Moments}

In this section, we introduce an approach to derive bounds on a network
spectral properties from its sequence of spectral moments. Since we have
expressions for the spectral moments in terms of structural properties,
these bounds relate the eigenvalues of a network with its structural
properties. For this purpose, we adapt an optimization framework proposed in 
\cite{Las02} and \cite{PB05} to derive optimal probabilistic bounds on a
random variable from a sequence of moments of its probability distribution.
In order to use this framework, we first need to introduce a probabilistic
interpretation of a network eigenvalue spectrum and its spectral moments.

For a simple graph $\mathcal{G}$, we define its spectral density as,%
\begin{equation}
\mu _{\mathcal{G}}\left( x\right) =\frac{1}{n}\sum_{i=1}^{n}\delta \left(
x-\lambda _{i}\right) ,  \label{Spectral Measure}
\end{equation}%
where $\delta \left( \cdot \right) $ is the Dirac delta function and $%
\left\{ \lambda _{i}\right\} _{i=1}^{n}$ is the set of (real) eigenvalues of
the symmetric adjacency matrix $A_{\mathcal{G}}$. Let us define a random
variable $X$ with probability density $\mu _{\mathcal{G}}$. The moments of $%
X\sim \mu _{\mathcal{G}}$ are equal to the spectral moments of $A_{\mathcal{G%
}}$, i.e.,%
\begin{align*}
\mathbb{E}_{\mu _{\mathcal{G}}}\left( X^{k}\right) & =\int_{\mathbb{R}%
}x^{k}\mu _{\mathcal{G}}\left( x\right) dx \\
& =\frac{1}{n}\sum_{i=1}^{n}\int_{\mathbb{R}}x^{k}\delta \left( x-\lambda
_{i}\right) dx \\
& =\frac{1}{n}\sum_{i=1}^{n}\lambda _{i}^{k}=m_{k}\left( A_{\mathcal{G}%
}\right) ,
\end{align*}%
for all $k\geq 0$. Furthermore, for a given Borel measurable set $T$, we have%
\begin{equation*}
\Pr \left( X\in T\right) =\int_{x\in T}\mu _{\mathcal{G}}\left( x\right) dx=%
\frac{1}{n}\left\vert \left\{ \lambda _{i}:\lambda _{i}\in T\right\}
\right\vert .
\end{equation*}%
In other words, the probability of the random variable $X$ being in a set $T$
is proportional to the number of eigenvalues of $A_{\mathcal{G}}$ in $T$.

In this probabilistic context, we can study two problems that are relevant
for the network dynamical behavior. Given a truncated sequence of spectral
moments, we formulate these problems as follows: \medskip

\begin{problem}
\label{Popescu problem}\emph{Find optimal bounds on the number of
eigenvalues that can lie in a given interval} $T$.
\end{problem}

\medskip

\begin{problem}
\label{Lasserre problem}\emph{Find bounds on the smallest and largest
eigenvalues of }$A_{\mathcal{G}}$\emph{.}
\end{problem}

\medskip

In the following subsections, we provide solutions to each one of the above
problems, from only the knowledge of a truncated sequence of moments.

\subsection{\label{Popescu Technique}Solution to Problem \protect\ref%
{Popescu problem}}

Our solution is based on the following classical problem in analysis:

\begin{problem}[Moment Problem]
\emph{Given a sequence of moments }$\left( m_{0},...,m_{k}\right) $\emph{,
and Borel measurable sets }$T\subseteq \Omega \subseteq \mathbb{R}$\emph{,
compute:}%
\begin{equation}
\begin{array}{lrll}
Z_{P}= & \max_{\mu } & \int_{T}1\,d\mu &  \\ 
& \text{s.t.} & \int_{\Omega }x^{j}\,d\mu =m_{j}, & \text{for }j=0,1,...,k.%
\end{array}
\label{Primal}
\end{equation}%
\emph{where }$\mu \in \mathbb{M}\left( \Omega \right) $\emph{, }$\mathbb{M}%
\left( \Omega \right) $\emph{\ being the set of positive Borel measures
supported by }$\Omega $\emph{.}
\end{problem}


The solution to this problem provides an extension to the classical Markov
and Chebyshev's inequalities in probability theory when moments of order
greater than 2 are available. In \cite{Las02} and \cite{PB05}, it was shown
that the optimal value of $Z_{P}$ can be efficiently computed by solving a
single semidefinite program using a dual formulation. Before we introduce
this dual formulation, it is important to discuss some details regarding the
feasibility of this problem.

A sequence of moments $\mathbf{m}_{k}=\left( m_{0},m_{1},...,m_{k}\right) $
is said to be feasible in $\Omega$ if there exists a measure $\mu\in \mathbb{%
M}\left( \Omega\right) $ whose moments match those in the sequence $\mathbf{m%
}_{k}$.\footnote{%
In what follows, we assume that our measures are densities, hence $m_{0}=1$.}
In general, an arbitrary sequence of numbers may not correspond to a
feasible sequence of moments. The problem of deciding whether or not a
sequence of numbers is a feasible sequence of moments is called the
classical moment problem \cite{ST43}. Depending on the set $\Omega $, we
find three important instances of this problem:

(\emph{i}) the Hamburguer moment problem, when $\Omega=\mathbb{R}$,

(\emph{ii}) the Stieltjes moment problem, when $\Omega=\mathbb{R}_{+}$, and

(\emph{iii}) the Hausdorff moment problem, when $\Omega =\left[ 0,1\right] $.

For univariate distributions, necessary and sufficient conditions for
feasibility of these instances of the classical moment problem can be given
in terms of certain matrices being positive semidefinite, as follows. Let us
define, for any $s\geq 0$, the following Hankel matrices of moments,%
\begin{align}
R_{2s}& =\left[ 
\begin{array}{cccc}
m_{0} & m_{1} & \cdots & m_{s} \\ 
m_{1} & m_{2} & \cdots & m_{s+1} \\ 
\vdots & \vdots & \ddots & \vdots \\ 
m_{s} & m_{s+1} & \cdots & m_{2s}%
\end{array}%
\right] ,  \label{Hankel matrices} \\
R_{2s+1}& =\left[ 
\begin{array}{cccc}
m_{1} & m_{2} & \cdots & m_{s+1} \\ 
m_{2} & m_{3} & \cdots & m_{s+2} \\ 
\vdots & \vdots & \ddots & \vdots \\ 
m_{s+1} & m_{s+2} & \cdots & m_{2s+1}%
\end{array}%
\right] .  \notag
\end{align}%
Then, we have the following feasibility results for the Hamburguer moment
problem \cite{ST43}:\footnote{%
Feasibility conditions for the Stieltjes and Hausdorff moment problems can
also be found in \cite{ST43}, but they are not relevant in this paper.}

\medskip

\begin{theorem}
\label{Hamburguer Feasibility}A necessary and sufficient condition for a
sequence of moments $\mathbf{m}_{k}=\left( m_{0},m_{1},...,m_{2s}\right) $
to be feasible in $\Omega=\mathbb{R}$ is $R_{2s}\succeq0$.
\end{theorem}

\medskip

Notice that the Hankel matrix associated to the sequence of spectral moments 
$\left( 1,m_{1}\left( A_{\mathcal{G}}\right) ,...,m_{2s}\left( A_{\mathcal{G}%
}\right) \right) $ of a finite graph $\mathcal{G}$ always satisfy Hamburguer
feasibility condition, $R_{2s}\succeq0$. We now describe the dual
formulation proposed in \cite{Las02} and \cite{PB05} to compute the solution
of the infinite-dimensional optimization problem in (\ref{Primal}) by
solving a single semidefinite program.

Using duality theory, one can associate a dual variable $y_{i}$ to each
equality constraint of the primal (\ref{Primal}) to obtain (see \cite{PB05}
for more details): 
\begin{equation}
\begin{array}{lrll}
Z_{D}= & \min_{y_{i}} & \sum_{i=0}^{k}y_{i}m_{i} &  \\ 
& \text{s.t.} & \sum_{i=0}^{k}y_{i}x^{i}-1\geq 0\text{,} & \text{for }x\in T,
\\ 
&  & \sum_{i=0}^{k}y_{i}x^{i}\geq 0\text{,} & \text{for }x\in \Omega .%
\end{array}
\label{Dual}
\end{equation}%
Notice that the dual constrains are univariate polynomials in $x$. Since a
univariate polynomial is nonnegative if and only if it can be written as sum
of squares of polynomials, the dual problem can be formulated as a
sum-of-squares program (SOSP) that can be numerically solved via
semidefinite programming. (For more details on SOSP and SDP, the interested
reader is referred to \cite{PPSP04} and \cite{VB96}.) Karlin and Isii proved
the following result concerning strong duality \cite{KS66}:

\medskip

\begin{theorem}
\label{Strong Duality}If the Hankel matrix of moments $R_{2s}$ defined in (%
\ref{Hankel matrices}) is positive definite, then $Z_{P}=Z_{D}$.
\end{theorem}

\medskip

From Theorem \ref{Hamburguer Feasibility}, we have that a sequence of
spectral moments satisfy strong duality if $\det R_{2s}\neq 0$. This
determinant is zero only for very degenerate networks, and we assume strong
duality holds for the networks studied in this paper.

\begin{figure}[t]
\centering
\includegraphics[width=0.45\textwidth]{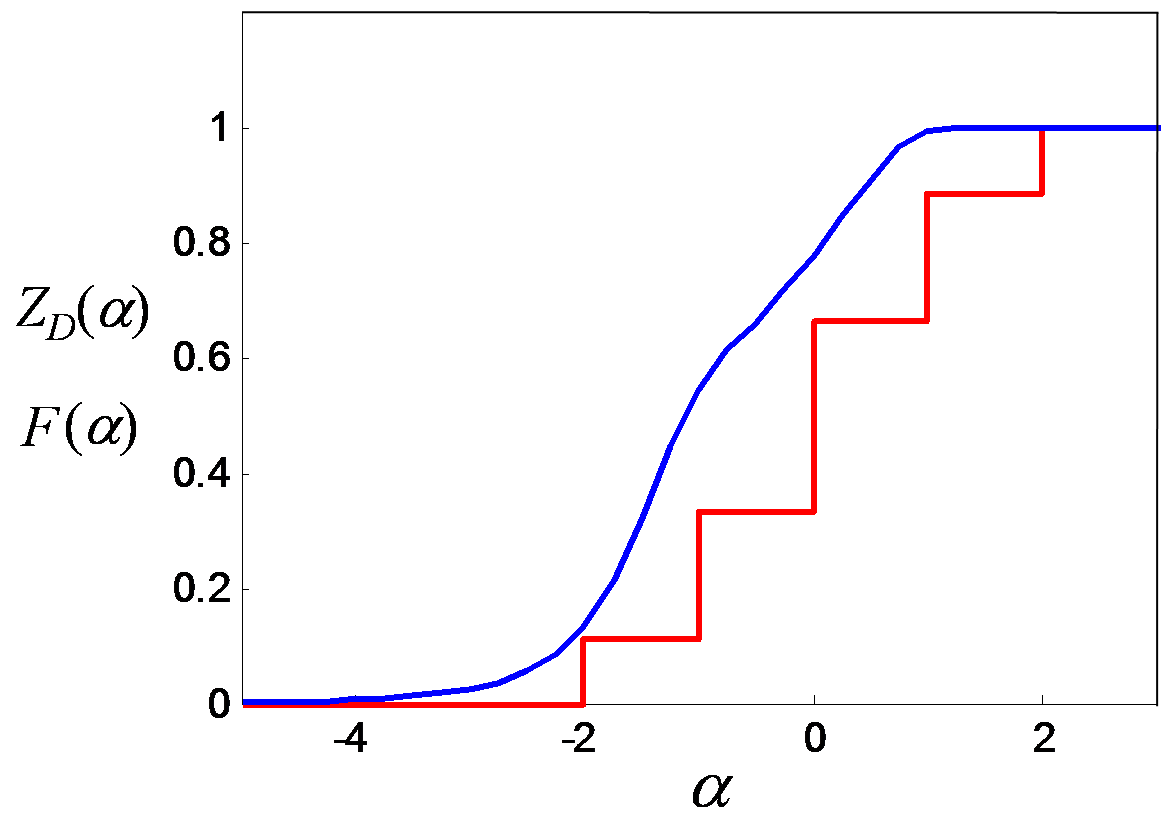}
\caption{Numerical solution of the SOSP described in Example \protect\ref%
{SOSP Example}. The stairs-like function corresponds to the cumulative
density function of $\protect\mu_{\mathcal{G}} $, $F(\protect\alpha)$.
Observe how the (numerical) function $Z_{D}\left( \protect\alpha\right) $ is
greater (or equal) to $F(\protect\alpha)$ for all values of $\protect\alpha$%
. }
\label{fig_8}
\end{figure}

The optimization framework we have described above can be used to solve
Problem \ref{Popescu problem}, since the solution of the dual problem in (%
\ref{Dual}), when $m_{j}=m_{j}\left( A_{\mathcal{G}}\right) $ for $%
j=1,2,...,k$, satisfies:%
\begin{equation}
Z_{D}\geq \int_{T}1~d\mu _{\mathcal{G}}=\frac{1}{n}\left\vert \left\{
\lambda _{i}:\lambda _{i}\in T\right\} \right\vert ,
\label{ZD is upper bound}
\end{equation}%
where $\mu _{\mathcal{G}}$ is the spectral density of $\mathcal{G}$. Then, $%
Z_{D}$ is the optimal upper bound on the number of eigenvalues of $A_{%
\mathcal{G}}$ that lie in the set $T$ given a truncated sequence of spectral
moments. We illustrate this result in the following example.

\begin{example}
\label{SOSP Example}Let us consider the spectral distribution $\mu _{%
\mathcal{G}}=\frac{1}{n}\sum_{i}w_{i}\delta \left( x-x_{i}\right) $, with
atomic masses located at $\left( x_{i}\right) _{1\leq i\leq 5}=\left(
-2,-1,0,1,2\right) $ and weights $\left( w_{i}\right) _{1\leq i\leq
5}=\left( 1/9,2/9,3/9,2/9,1/9\right) $. The sequence of moments of $\mu _{%
\mathcal{G}}$ is $\left( m_{k}\right) _{1\leq i\leq 5}=\left(
0,4/3,0,4,0\right) $. Let us define $F\left( \alpha \right) \triangleq
\int_{-\infty }^{\alpha }\mu _{\mathcal{G}}\left( x\right) dx$, i.e., the
cumulative distribution of $\mu _{\mathcal{G}}\left( x\right) $, and denote
by $Z_{D}\left( \alpha \right) $ the numerical solution to the dual SOSP in (%
\ref{Dual}) for $T=\left( -\infty ,\alpha \right] $. According to (\ref{ZD
is upper bound}), $Z_{D}\left( \alpha \right) $ is an upper bound of $%
F\left( \alpha \right) $ for all values of $\alpha $. In Fig. 4, we verify
this result by plotting the cumulative distribution $F\left( \alpha \right) $
and $Z_{D}\left( \alpha \right) $ for $\alpha =\left[ -5:0.25:3\right] $.
\end{example}


\subsection{\label{Solution Lasserre Problem}Solution to Problem \protect\ref%
{Lasserre problem}}

In this subsection, we derive bounds on the smallest and largest eigenvalues
of $A_{\mathcal{G}}$ from only the knowledge of a truncated sequence of
spectral moments. For this purpose, we apply the technique proposed in \cite%
{Las11} to compute the smallest interval $\left[ a,b\right] $ containing the
support\footnote{%
Recall that the support of a finite Borel measure $\mu $ on $R$, denoted by $%
supp\left( \mu \right) $, is the smallest closed set $B$ such that $\mu
\left( R\backslash B\right) =0$.} of a positive Borel measure $\mu $ from
its complete sequence of moments $\left( m_{r}\right) _{r\geq 0}$. In \cite%
{Las11} a technique was also proposed to compute tight bounds on the values
of $a$ and $b$ when only a \emph{truncated} sequence of moments $\left(
m_{r}\right) _{0\leq r\leq k}$ is known. In the context of spectral graph
theory, we can apply this technique to a sequence of spectral moments in
order to bound the support of the spectral measure $\mu _{\mathcal{G}}$ of a
graph $\mathcal{G}$. In this context, the extreme values, $a$ and $b$, of
the smallest interval containing the support of $\mu _{\mathcal{G}}$
corresponds to the minimum and maximum eigenvalues of $A_{\mathcal{G}}$,
denoted by $\lambda _{\min }\left( A_{\mathcal{G}}\right) $ and $\rho \left(
A_{\mathcal{G}}\right) $, respectively. Since we can compute the first five
spectral moments in terms of structural properties using the results in
Section \ref{Moments from Metrics}, this technique allows to compute bounds
on $\lambda _{\min }\left( A_{\mathcal{G}}\right) $ and $\rho \left( A_{%
\mathcal{G}}\right) $ in terms of structural properties.

We describe the scheme proposed in \cite{Las11} to compute the smallest
interval $\left[ a,b\right] $ by solving a series of SDPs in one variable.
As we shall show below, at step $s$ of this series of SDPs, we are given a
sequence of moments $\left( m_{1},...,m_{2s+1}\right) $ and solve two SDPs
whose solution provides an inner approximation $\left[ \alpha _{s},\beta _{s}%
\right] \subseteq \left[ a,b\right] $. As we increase $s$ in this series, we
obtain two sequences $\left( \alpha _{s}\right) _{s\in \mathbb{N}}$ and $%
\left( \beta _{s}\right) _{s\in \mathbb{N}}$ that are respectively monotone
nonincreasing and nondecreasing, and converge to $a$ and $b$ as $%
s\rightarrow \infty $. In our case, we have expressions for the first five
spectral moments, $\left( m_{1},...,m_{5}\right) $, hence, we can solve the
first two steps of the series of SDPs. The solutions, $\alpha _{s}$ and $%
\beta _{s}$, of these SDPs provide upper and lower bounds on $\lambda _{\min
}\left( A_{\mathcal{G}}\right) $ and $\rho \left( A_{\mathcal{G}}\right) $,
respectively, in terms of structural properties.

In order to formulate the series of SDPs proposed in \cite{Las11} we need to
define the so-called localizing matrix \cite{LasBOOK}. Given a sequence of
moments, $\mathbf{m}^{\left( 2s+1\right) }=\left( m_{1},...,m_{2s+1}\right) $%
, the localizing matrix is a Hankel matrix defined as:%
\begin{equation}
H_{s}\left( c\right) \triangleq R_{2s+1}-c~R_{2s}\text{,}
\label{Localizing matrix}
\end{equation}%
where $R_{2s}$ and $R_{2s+1}$ are the Hankel matrices of moments defined in (%
\ref{Hankel matrices}). Hence, for a given sequence of moments, the entries
of $H_{s}\left( c\right) $ depend affinely on the variable $c$. Then, we can
compute $\alpha _{s}$ and $\beta _{s}$ as follows \cite{Las11}:

\begin{proposition}
\label{Lasserre bound}Let $\mathbf{m}^{\left( 2s+1\right) }=\left(
m_{1},...,m_{2s+1}\right) $ be the truncated sequence of moments of a
positive Borel measure $\mu $. Then,%
\begin{align}
a& \leq \alpha _{s}\triangleq \max_{\alpha }\left\{ \alpha :H_{s}\left(
\alpha \right) \succcurlyeq 0\right\} ,  \label{Bound min eigenval} \\
b& \geq \beta _{s}\triangleq \min_{\beta }\left\{ \beta :-H_{s}\left( \beta
\right) \succcurlyeq 0\right\} ,  \label{Bound max eigenval}
\end{align}%
for $\left[ a,b\right] $ being the smallest interval containing $supp(\mu )$.
\end{proposition}

\medskip

\begin{remark}
Observe that $\alpha _{s}$ and $\beta _{s}$ are the solutions to two SDPs in
one variable, which can be efficiently solved using standard
optimization software (for example, CVX \cite{CVX}). Notice that the matrix
involved in the semidefinite constrains in (\ref{Bound min eigenval}) and (%
\ref{Bound max eigenval}), $H_{s}\left( x\right) $, has size $(s+1)\times
(s+1)$. Hence, the computational complexity of solving this SDP is
polynomial in $s$, \cite{Las11}. Since $s$ is a small number in our context
(i.e., $s=2$ if we use five moments in Proposition 1), the
computational cost of solving this SDP is negligible in comparison with the
cost of counting triangles, quadrangles and pentagons in $\mathcal{G}$,
which requires $\sum_{i=1}^{n}\binom{N_{i,1}}{2}$, $\sum_{i=1}^{n}\binom{N_{i,2}%
}{3}$, and $\sum_{i=1}^{n}\binom{N_{i,2}}{4}$ operations, respectively.
\end{remark}

Given a truncated sequence of spectral moments, the smallest interval $\left[
a,b\right] $ becomes $\left[ \lambda _{\min }\left( A_{\mathcal{G}}\right)
,\rho \left( A_{\mathcal{G}}\right) \right] $, thus, Proposition \ref%
{Lasserre bound}\ provide an efficient numerical scheme to compute the
bounds $\alpha _{s}\geq \lambda _{\min }\left( A_{\mathcal{G}}\right) $ and $%
\beta _{s}\leq \rho \left( A_{\mathcal{G}}\right) $. Furthermore, for $s=1$
and $2$, one can analytically solve the SDPs in (\ref{Bound min eigenval})
and (\ref{Bound max eigenval}). For example, in the case $s=1$, we are given
a sequence of three spectral moments, $\left( m_{1},m_{2},m_{3}\right) $,
with localizing matrix%
\begin{equation*}
H_{1}\left( c\right) =\left[ 
\begin{array}{cc}
m_{1}-cm_{0} & m_{2}-cm_{1} \\ 
m_{2}-cm_{1} & m_{3}-cm_{2}%
\end{array}%
\right] .
\end{equation*}%
From (\ref{Moments as Averages}), the spectral moments of simple graphs
satisfy $m_{0}=1$, $m_{1}=0$, $m_{2}=2e/n$, and $m_{3}=6\Delta /n$. One can
prove that the optimal values of $\alpha _{1}$ and $\beta _{1}$ are the
smallest and largest root of $\det H_{1}(c)=0$, which is a second-order
polynomial in the variable $c$. Then, we have the following bounds on $\rho
\left( A_{\mathcal{G}}\right) $ and $\lambda _{\min }\left( A_{\mathcal{G}%
}\right) $ in terms of the number of nodes $n$, edges $e$, and triangles $%
\Delta $ in $\mathcal{G}$: 
\begin{align}
\rho \left( A_{\mathcal{G}}\right) & \geq \beta _{1}=\frac{6\Delta +\sqrt{%
36\Delta ^{2}+32e^{3}/n}}{4e},  \label{Lasserre bound 3 moments} \\
\lambda _{\min }\left( A_{\mathcal{G}}\right) & \leq \alpha _{1}=\frac{%
6\Delta -\sqrt{36\Delta ^{2}+32e^{3}/n}}{4e}.  \notag
\end{align}

As mentioned above, tighter bounds on the spectral radius can be found as we
increase the value of $s$ in Proposition \ref{Lasserre bound}. In the case $%
s=2$, we are given a sequence of five spectral moments $\left(
m_{1},m_{2},...,m_{5}\right) $ with localizing matrix,%
\begin{equation}
H_{2}\left( c\right) =\left[ 
\begin{array}{ccc}
m_{1}-c & m_{2}-cm_{1} & m_{3}-cm_{2} \\ 
m_{2}-cm_{1} & m_{3}-cm_{2} & m_{4}-cm_{3} \\ 
m_{3}-cm_{2} & m_{4}-cm_{3} & m_{5}-cm_{4}%
\end{array}%
\right] .  \label{H2c}
\end{equation}%
As we proved in Section \ref{Moments from Metrics}, these moments depend on
the number of nodes, edges, cycles of length 3 to 5, the sum of squares of
degrees $W_{2}$, and the degree-triangle correlation $\mathcal{C}_{dt}$.
Since we are using much richer structural information than in the case $s=1$%
, we should expect the resulting bounds to be substantially tighter (as we
shall verify in Section \ref{Simulations}). For $s=2$, the optimal values $%
\alpha _{2}$ and $\beta _{2}$ can also be analytically computed, as follows.
First, note that $-H_{2}(c)\succcurlyeq 0$ if and only if all the
eigenvalues of $H_{2}$ are nonpositive. The characteristic polynomial of $%
H_{2}\left( c\right) $ can be written as%
\begin{equation*}
\phi _{2}\left( \lambda \right) \triangleq \det \left( \lambda
I-H_{2}(c)\right) =\lambda ^{3}+p_{1}\lambda ^{2}+p_{2}\lambda +p_{3},
\end{equation*}%
where $p_{j}$ is a polynomial of degree $j$ in the variable $c$ (with
coefficients depending on the moments). Thus, by Descartes' rule, all the
eigenvalues of $H_{2}$ are nonpositive if and only if $p_{j}\geq 0$, for $%
j=1,2,$ and $3$. In fact, one can prove that the optimal values of $\alpha
_{2}$ and $\beta _{2}$ in (\ref{Bound min eigenval}) and (\ref{Bound max
eigenval}) can be computed as the smallest and largest roots of $p_{3}\left(
c\right) =\det H_{2}\left( c\right) =0$, which is a third degree polynomial
in the variable $c$ \cite{Las11}. Therefore, the expressions that allow us
to compute the optimal bounds are:%
\begin{align}
\rho \left( A_{\mathcal{G}}\right) & \geq \beta _{2}=\max \{\text{roots}%
[p_{3}(c)]\},  \label{Optimal Fifth Bound} \\
\lambda _{\min }\left( A_{\mathcal{G}}\right) & \leq \alpha _{2}=\min \{%
\text{roots}[p_{3}(c)]\},  \notag
\end{align}%
where $p_{3}\left( c\right) =d_{3}c^{3}+d_{2}c^{2}+d_{1}c+d_{0,}$ with%
\begin{align*}
d_{0}& \triangleq
2m_{2}m_{3}m_{4}-m_{5}m_{2}^{2}-m_{3}^{3}+m_{1}m_{5}m_{3}-m_{1}m_{4}^{2}, \\
d_{1}& \triangleq
m_{2}m_{3}^{2}-m_{2}^{2}m_{4}+m_{1}m_{5}m_{2}-m_{1}m_{3}m_{4} \\
& -m_{5}m_{3}+m_{4}^{2}, \\
d_{2}& \triangleq
m_{4}m_{1}m_{2}-m_{5}m_{1}^{2}+m_{1}m_{3}^{2}-m_{2}^{2}m_{3} \\
& +m_{5}m_{2}-m_{4}m_{3}, \\
d_{3}& \triangleq
m_{4}m_{1}^{2}-2m_{1}m_{2}m_{3}+m_{2}^{3}-m_{4}m_{2}+m_{3}^{2}.
\end{align*}%
There are closed-form expressions for the roots of this third-order
polynomial (for example, Cardano's formula \cite{AS65}), although the
resulting expressions for the roots are rather complicated and do not
provide much insight.

\begin{figure}[t]
\centering
\includegraphics[width=0.43\textwidth]{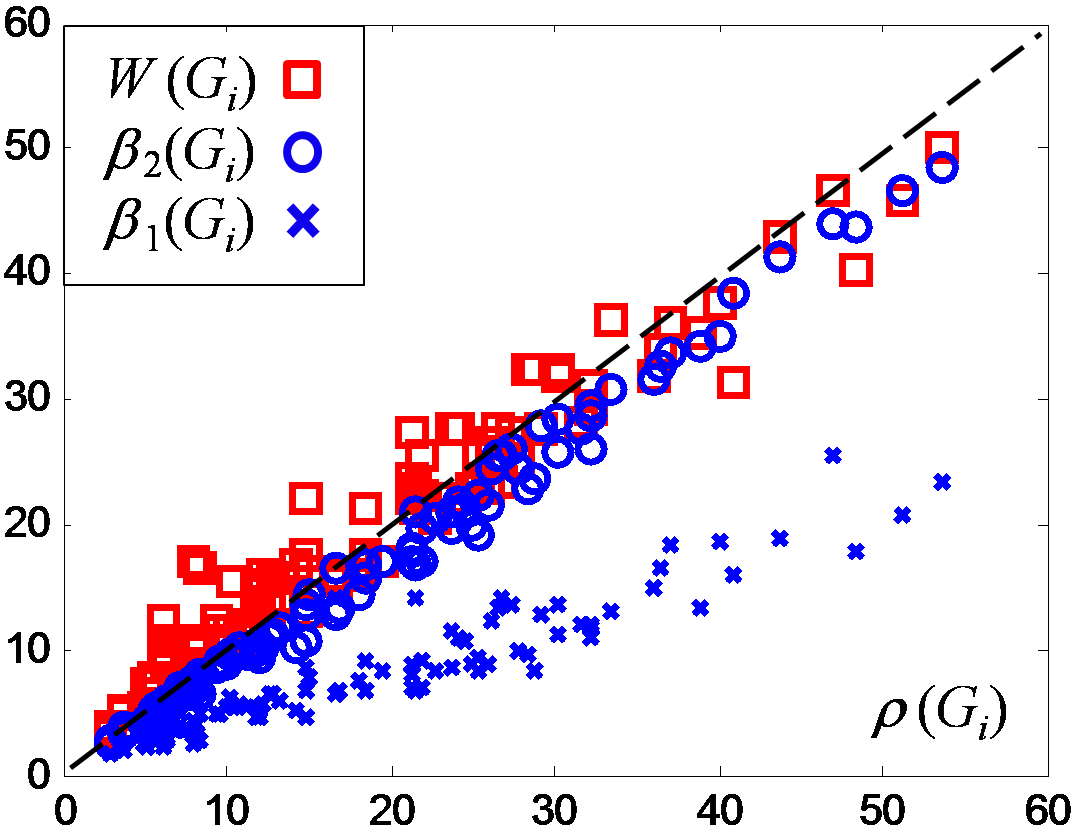}
\caption{Scatter plot of the spectral radius, $\protect\rho \left(
G_{i}\right) $, versus the lower bounds $\protect\beta _{1}\left(
G_{i}\right) $ (crosses) and $\protect\beta _{2}\left( G_{i}\right) $
(circles), as well as the random--graph-based estimator $W\left(
G_{i}\right) $ (squares), where each point is associated with one of the $%
100 $ social subgraphs considered in our experiments.}
\end{figure}

In this subsection, we have presented a convex optimization framework to
compute optimal bounds on the maximum and minimum eigenvalues of a graph $%
\mathcal{G}$ from a truncated sequence of its spectral moments. Since we
have expressions for spectral moments in terms of structural properties,
these bounds relate the eigenvalues of a graph with its structural
properties.

\section{\label{Simulations}Numerical Simulations and Structural Implications%
}

In this section, we analyze real data from a regional network of Facebook
that spans $63,731$ users (nodes) connected by $817,090$ friendships (edges) 
\cite{VMCG09}. In order to corroborate our results in different network
topologies, we extract multiple medium-size social subgraphs from the
Facebook graph by running a Breath-First Search (BFS) around different
starting nodes. Each BFS induces a social subgraph spanning all nodes 2 hops
away from a starting node. We use this approach to generate a set $\mathbf{G}%
=\{G_{i}\}_{i\leq 100}$ of 100 different social subgraphs centered around
100 randomly chosen nodes.\footnote{%
Although this procedure is common in studying large social network , it
introduces biases that must be considered carefully \cite{SWM05}.}



\begin{figure}[t]
\centering
\includegraphics[width=0.43\textwidth]{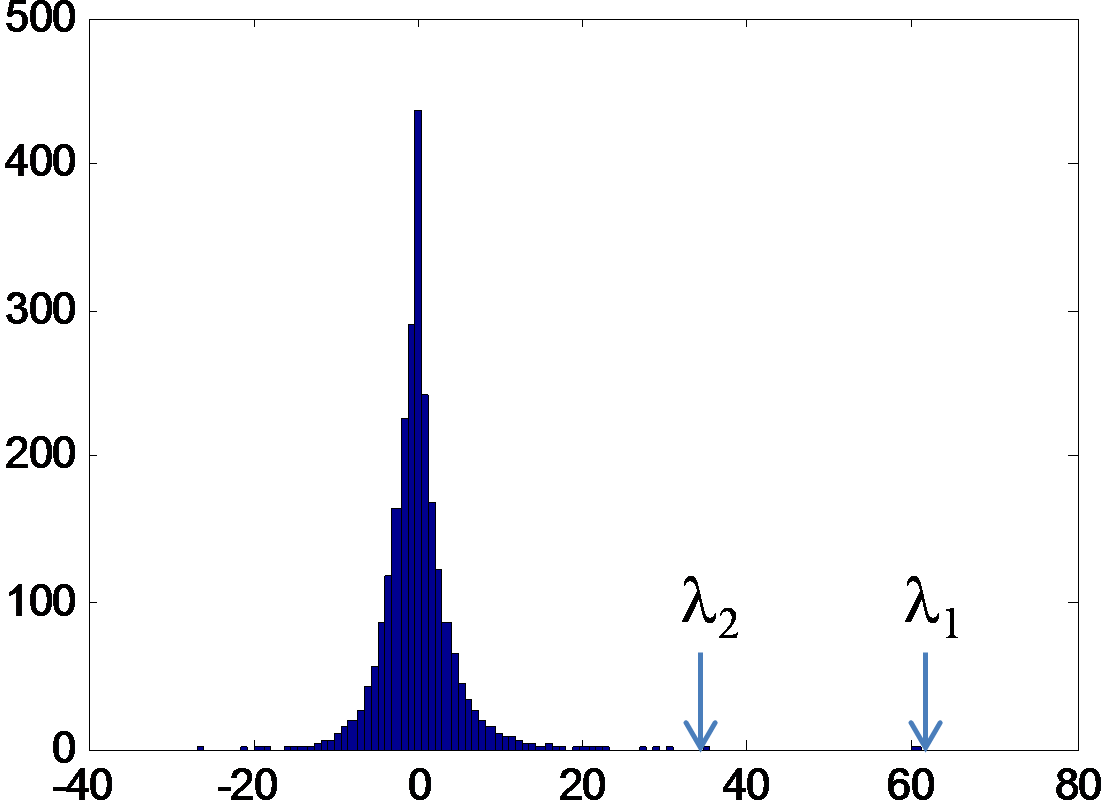}
\caption{Histogram of eigenvalues of a social subgraph of Facebook with
2,404 nodes.}
\end{figure}

In our first numerical experiment, we compute the first five spectral
moments $\mathbf{m}_{5}\left( G_{i}\right) =\left( m_{1}\left( G_{i}\right)
,...,m_{5}\left( G_{i}\right) \right) $ for each social subgraph $G_{i}\in 
\mathbf{G}$. From these moments, we then compute the lower bounds on the
spectral radius $\beta _{1}\left( G_{i}\right) $ and $\beta _{2}\left(
G_{i}\right) $ using Proposition \ref{Lasserre bound}. Fig. 5 is a scatter
plot where each cross has coordinates $\left( \rho \left( G_{i}\right)
,\beta _{1}\left( G_{i}\right) \right) $ and each circle has coordinates $%
\left( \rho \left( G_{i}\right) ,\beta _{2}\left( G_{i}\right) \right) $,
for all $G_{i}\in \mathbf{G}$. In the same figure, we have also included a
cloud of red squares with coordinates $\left( \rho \left( G_{i}\right)
,W\left( G_{i}\right) \right) $, where $W\left( G_{i}\right) \triangleq
\left. \sum_{i}d_{i}^{2}\right/ \sum_{i}d_{i}$ is the estimator based on
synthetic random graphs (\ref{CL Estimator}). Observe how 
\emph{the spectral radii }$\rho \left( G_{i}\right) $ \emph{of these social
subgraphs are remarkably close to the theoretical lower bound} $\beta
_{2}\left( G_{i}\right) $. In particular, the correlation coefficient
between $\rho \left( G_{i}\right) $ and $\beta _{2}\left( G_{i}\right) $ is
equal to 0.995 (while the correlation between $\rho \left( G_{i}\right) $
and the estimator based on random networks, $W\left( G_{i}\right) $, is
equal to 0.974). Therefore, it is reasonable to use $\beta _{2}\left(
G_{i}\right) $ as an estimate of $\rho \left( G_{i}\right) $ for social
subgraphs.

In what follows, we analyze the spectral moments of online social networks
to reveal the set of structural properties having the highest impact on the
spectral radius. Empirical evidence strongly suggests that many real-world
networks present heavy-tailed eigenvalue distributions \cite{FDBV01},\cite%
{MP02}. For example, in Fig. 6 we have included the histogram of eigenvalues
of a subgraph of Facebook with $2,404$ nodes, where we can observe the
following two typical properties in the spectrum of online social networks: (%
$i$)\ The largest eigenvalue of the network is well separated from the rest
of eigenvalues (spectral dominance), and ($ii$) the bulk of eigenvalues
concentrates around the origin. In this case, we can numerically approximate
high-order moments as $m_{k}\left( A_{\mathcal{G}}\right) \approx \frac{1}{n}%
\lambda _{1}^{k}$. Furthermore, we have from (\ref{Fifth Moment Subgraphs})
that the fifth spectral moment is equal to $m_{5}\left( A_{\mathcal{G}%
}\right) =\frac{1}{n}\left[ 10\Pi +10\mathcal{C}_{dt}-30\Delta \right]
\approx \frac{1}{n}\lambda _{1}^{5}$. Therefore, we have the following
estimator for the spectral radius of online social networks:%
\begin{equation}
\lambda _{1}\approx \tilde{\lambda}_{1}^{\left( a\right) }\triangleq \left(
10\Pi +10\mathcal{C}_{dt}-30\Delta \right) ^{1/5}\text{.}
\label{Spectral Radius Dominance}
\end{equation}
For example, for the social subgraph with 2,404 nodes mentioned above, the
exact value of the spectral radius is $\lambda _{1}=60.9$, while the
estimator is $\tilde{\lambda}_{1}^{(a)}=62.6$.

We now use (\ref{Spectral Radius Dominance}) to unveil the set of structural
properties that are most influential on the spectral radius. In Fig. 7, we
plot (in semilog scale) the values of $\Pi $, $\mathcal{C}_{dt}$, and $%
\Delta $ for each one of the 100 different social subgraphs, $G_{i}\in 
\mathbf{G}$, considered in our previous experiment. Observe how the number
of triangles $\Delta $ is always much smaller than $\Pi +\mathcal{C}_{dt}$.
Therefore, for online social networks, we can simplify (\ref{Spectral Radius
Dominance}) as follows,%
\begin{equation*}
\lambda _{1}\approx \tilde{\lambda}_{1}^{\left( b\right) }\triangleq \left(
10\Pi +10\mathcal{C}_{dt}\right) ^{1/5}.
\end{equation*}
Fig. 8 is a scatter plot where each circle has coordinates $(\rho \left(
G_{i}\right) ,\tilde{\lambda}_{1}^{\left( b\right) }\left( G_{i}\right) )$,
for all $G_{i}\in \mathbf{G}$. We observe how our approximation presents an
excellent performance in practice, outperforming the popular estimator, $%
W\left( G_{i}\right) $, based on random networks (see Fig. 5).

\begin{figure}[t]
\centering
\includegraphics[width=0.43\textwidth]{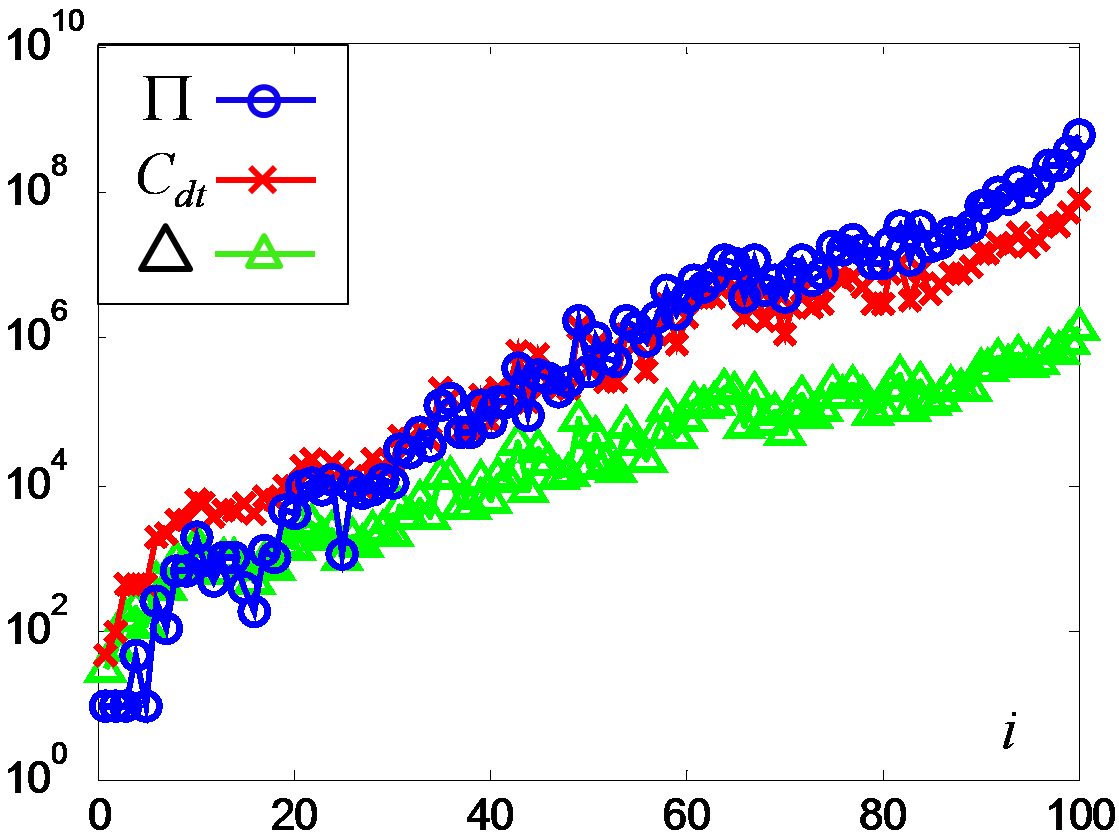}
\caption{Number of triangles $\Delta $ (green triangles), degree-triangle
correlation $C_{dt}$ (red crosses), and number of pentagons $\Pi $ (blue
circles) for each one of the $100$ social subgraphs considered in our
experiments (in semilog scale).}
\end{figure}

The estimator $\tilde{\lambda}_{1}^{\left( b\right) }$ provides a clear
insight about what structural properties have the strongest impact in the
spectral radius of online social networks. In particular, $\tilde{\lambda}%
_{1}^{\left(b\right) }$ unveils that both the number of pentagons, $\Pi $,
and the degree-triangle correlation, $\mathcal{C}_{dt}$, are structural
properties with a strong influence on the spectral radius.

\bigskip 

The tightness of our bounds depends on the nature of the data used. In
the following examples, we illustrate the quality of our bounds for an
Internet and an e-mail network:

\medskip

\begin{example}[Enron e-mail network]
In this example we study the spectral properties of a subgraph of the Enron
e-mail network \cite{BY04}. In this network, nodes correspond to e-mail
addresses and an edge $\left( i,j\right) $ exists if $i$ sent at least one
e-mail to $j$ (or vice versa). The subgraph under study has $n=3,215$ nodes, 
$e=36,537$ edges, and its largest eigenvalue is $\lambda _{1}=95.18$. Using
the results in Section \ref{Moments from Metrics}, we compute the first five
spectral moments of the adjacency matrix to be: $m_{1}=0$, $m_{2}=22.47$, $%
m_{3}=394.7$, $m_{4}=33,491$, and $m_{5}=2,603,200$. From Proposition \ref%
{Lasserre bound}, we obtain the following lower bound on the largest
eigenvalue: $\beta _{2}=78.53<\lambda _{1}$. We can also compare our bound
with the estimator in (\ref{CL Estimator}), corresponding to
a random network with the same degree distribution. The value of the
estimator is equal to $\tilde{\lambda}_{1}=124.57$.
\end{example}

\medskip

\begin{example}[AS-Skitter Internet network]
We now consider a subgraph of the Internet network at the Autonomous Systems
(AS) level, which was obtained from the Skitter data collection in CAIDA 
\cite{CAIDA05}. Our subgraph has $n=2,248$ nodes, $e=20,648$ edges, and its
largest eigenvalue at $\lambda _{1}=91.3$. The spectral moments of its
adjacency matrix are $m_{1}=0$, $m_{2}=18.37$, $m_{3}=341.1$, $m_{4}=40,001$%
, and $m_{5}=2,777,018$, and the resulting lower bound is $\beta
_{2}=74.72<\lambda _{1}$. In this case, the estimator based on random
networks produces a value of $\tilde{\lambda}_{1}=219.1$, which is very
loose. Therefore, estimators based on random networks can be very misleading
in the analysis of the Internet graph.
\end{example}

In this section, we have first shown that $\beta _{2}\left( G_{i}\right) $
can be used as an estimator of the spectral radius $\rho \left( G_{i}\right) 
$ for online social subgraphs, outperforming the estimator based on random
networks. Furthermore, we have analyzed the spectral moments of online
social networks to unveil the set of structural properties having the
highest impact on the spectral radius. In particular, we have found that the
number of pentagons and the degree-triangle correlation strongly influence
the spectral radius of online social networks.

\begin{figure}[t]
\centering
\includegraphics[width=0.43\textwidth]{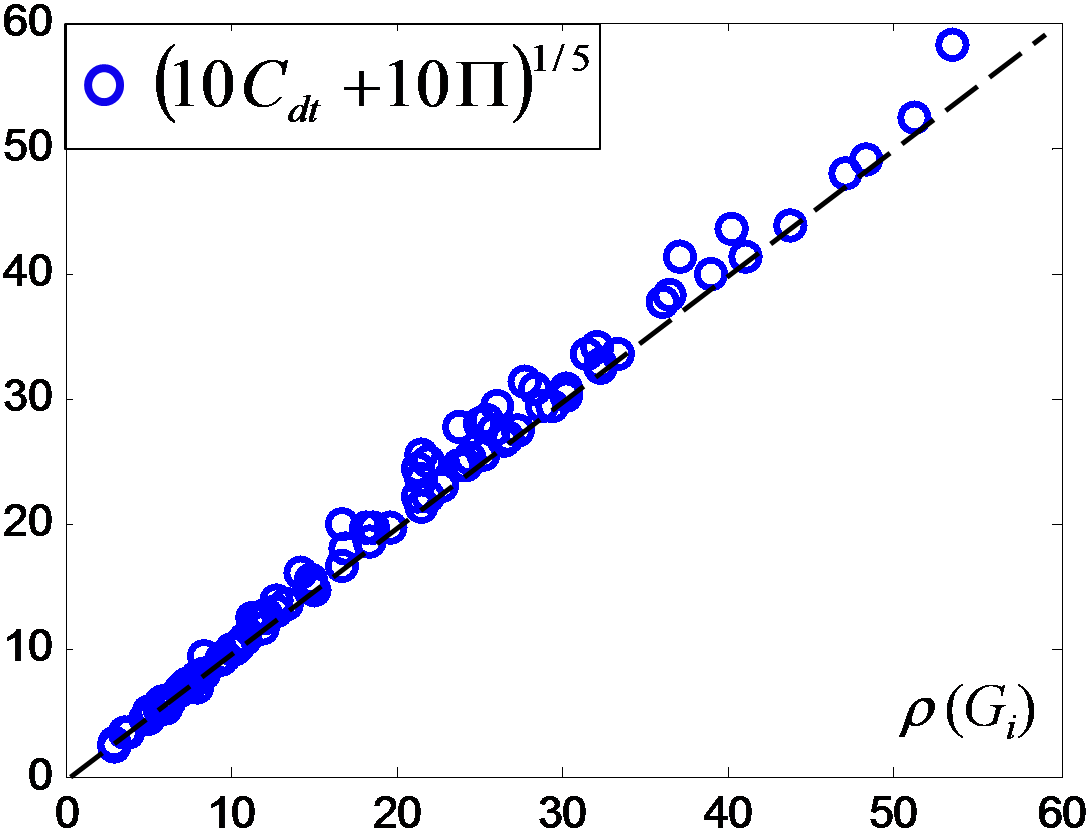}
\caption{Scatter plot of the spectral radius, $\protect\rho \left(
G_{i}\right) $, versus the spectral estimator $\tilde{\protect\lambda}%
_{i}^{\left( b\right) }\left( G_{i}\right) $, where each circle is
associated with one of the $100$ social subgraphs considered in our
experiments.}
\end{figure}

\section{Conclusions}

A fundamental question in the field of network science is to understand the
relationship between the structural properties of a network and its
dynamical performance. The common approach to study this relationship is to
use synthetic network models. Although very common, synthetic models present
some major flaws: (\emph{i}) These models are only suitable to study a very
limited range of structural properties, and (\emph{ii}) they implicitly
induce structural properties that are not directly controlled and can
influence the network dynamical performance.

In this paper, we have proposed an alternative approach to study the
relationship between a network structure and its dynamics that is not based
on synthetic models. Our approach exploits the closed connection between the
dynamical performance of many dynamical processes that can take place in a
network and its eigenvalue spectrum. Consequently, we have studied how
structural properties of a network relate to its eigenvalue spectrum using
algebraic graph theory and convex optimization. In particular, we have
derived expressions that explicitly relate structural properties of a
network with its spectral moments. We have also introduced an optimization
framework that allows us to extract optimal bounds on spectral properties of
interest using semidefinite programming.

Using our approach, we have unveiled those structural properties that have
the strongest impact on the spectral properties of a collection of social
subgraphs. In particular, we have found that the number of close cycles of
lengths 3 to 5 (quantified by $\Delta $, $Q$ and $\Pi $), as well as the sum
and sum-of-squares of the degrees, and the degree-triangle correlation $%
\mathcal{C}_{dt}$ have a direct influence on the eigenvalue spectrum.
Furthermore, in the case of online social networks, we have found that the
number of pentagons and the degree-triangle correlation strongly influence
the spectral radius of the network.

\appendix

\section{Proof of Lemma \protect\ref{Lemma 5th metrics}}

\textbf{Lemma \ref{Lemma 5th metrics} }Let $\mathcal{G}$ be a simple graph.
Denote by $p_{i}$, $t_{i}$, and $d_{i}$ the number of pentagons, triangles,
and edges touching node $i$ in $\mathcal{G}$, respectively. Then,%
\begin{equation*}
m_{5}\left( A_{\mathcal{G}}\right) =\frac{1}{n}\left[
\sum_{i=1}^{n}2p_{i}+10t_{i}d_{i}-10t_{i}\right] .
\end{equation*}

\begin{proof}
As in Lemma \ref{Lemma 4th metrics}, we count the number of closed walks of
length $5$ in $\mathcal{G}$. We classify these walks based on the structure
of the subgraph underlying each walk. We provide a classification of the
walk types in Fig. 2, where we also include expressions for the number of
closed walks of each type. We now provide the details on how to compute
those expressions for each walk type:

\begin{description}
\item[(a)] The number of closed walks of this type starting at $i$ is equal
to twice the number of pentagons touching $i$, hence, the total number is
given by $\sum_{i=1}^{n}2p_{i}$.

\item[(b)] In order to count walks of this type, it is convenient to define $%
t_{pqr}$ as the indicator function that takes value $1$ if there exists a
triangle connecting vertices $p$, $q$, and $r$ (0, otherwise). Note that
this indicator satisfies $\sum_{q=1}^{n}\sum_{r=1}^{n}t_{pqr}=2t_{p}$, where 
$t_{p}$ is the number of triangles touching node $p$. Hence, the number of
closed walks of type (b) can be written as:%
\begin{equation*}
w_{5}^{\left( b\right) }=\sum_{i=1}^{n}\sum_{p=1}^{n}\sum_{q\neq
i}\sum_{r\neq i}a_{ip}t_{pqr},
\end{equation*}%
where $a_{ip}$ indicates the existence of an edge from $i$ to $p$, and $%
t_{pqr}$ indicates the existence of a triangle connecting $q$ and $r$ with $%
p $. We can then perform the following algebraic manipulations,%
\begin{align*}
w_{5}^{\left( b\right) }& \overset{(i)}{=}\sum_{p=1}^{n}\sum_{q=1}^{n}%
\sum_{r=1}^{n}t_{pqr}\left( \sum_{i\neq q,r}a_{ip}\right) \\
& \overset{(ii)}{=}\sum_{p=1}^{n}\left( d_{p}-2\right)
\sum_{q=1}^{n}\sum_{r=1}^{n}t_{pqr} \\
& =2\sum_{p=1}^{n}t_{p}\left( d_{p}-2\right) ,
\end{align*}%
where in equality ($i$) we have changed the order of the subindices, and
impose the inequality constrains on subindex $i$. In equality ($ii$), we
take into account that $\sum_{i\neq q,r}a_{ip}=d_{i}-2$, since $p$ is
connected to $q$ and $r$ in this walk type.

\item[(c)] We can use the indicator function $t_{ijk}$ to write the total
number of walks in this type as follows,%
\begin{align*}
w_{5}^{\left( c\right) } &
=2\sum_{i=1}^{n}\sum_{j=1}^{n}\sum_{k=1}^{n}t_{ijk}\left( d_{j}-2\right) \\
& =4\sum_{j=1}^{n}\left( d_{j}-2\right) t_{j},
\end{align*}
where the last expression comes from reordering the summations and $\sum
_{i=1}^{n}\sum_{k=1}^{n}t_{ijk}=2t_{j}$.

\item[(d)] The number of walks starting at $i$ in this type is equal to $%
4t_{i}\left( d_{i}-2\right) $, where we have included a $-2$ in the
parenthesis to take into account that two of the edges touching $i$ are part
of the triangle. The coefficient $4$ accounts for the two possible direction
we can walk the triangle and the two possible choices for the first step of
the walk (towards the triangle or towards the single edge).

\item[(e)-(f)] These types of walks correspond to the set of closed walks of
length $5$ that visit all (and only) the edges of a triangle. Given a
particular triangle touching $i$, we can count the number of walks of this
type to be equal to $10$, where $8$ of them are of type (e) and $2$ of type
(f).
\end{description}

Hence, we obtain (\ref{Fifth Moment Metrics}) by summing up all the above
contributions (and simple algebraic manipulations).
\end{proof}

\begin{biography}
[{%
\includegraphics[width=1in,height=1.25in,clip,keepaspectratio]{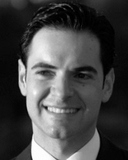}%
}] {Victor M. Preciado} received the Ph.D. degree in electrical engineering
and computer science from the Massachusetts Institute of Technology,
Cambridge, in 2008.

He is currently an Assistant Professor with the Department of Electrical
and Systems Engineering, University of Pennsylvania,
Philadelphia. His research interests lie in the modeling, analysis, and control of dynamical processes in large-scale complex networks, with applications in social technological and biological networks.
\end{biography}

\begin{biography}
[{%
\includegraphics[width=1in,height=1.25in,clip,keepaspectratio]{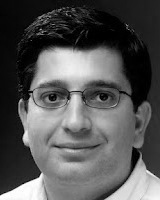}}]{%
Ali Jadbabaie} (SM'07) received the B.S. degree from Sharif University of
Technology, Teheran, Iran, in 1995, the M.S. degree in electrical and
computer engineering from the University of New Mexico, Albuquerque, in
1997, and the Ph.D. degree in control and dynamical systems from the
California Institute of Technology, Pasadena, in 2001.

From July 2001 to July 2002, he was a Postdoctoral Associate with the
Department of Electrical Engineering, Yale University, New Haven. Since July
2002, he has been with the Department of Electrical and Systems Engineering
and GRASP Laboratory, University of Pennsylvania, Philadelphia. His research is
broadly in control theory and network science, specifically, analysis,
design and optimization of networked dynamical systems with applications to
sensor networks, multi-vehicle control, social aggregation and other
collective phenomena.

Dr. Jadbabaie received the NSF Career Award, the ONR Young Investigator
Award, the Best Student Paper Award (as advisor) of the American Control
Conference 2007, the O. Hugo Schuck Best Paper Award of the American
Automatic Control Council, and the George S. Axelby Outstanding Paper Award
of the IEEE Control Systems Society.
\end{biography}

\end{document}